\documentclass[twocolumn,english]{autart}

\pdfoutput=1
\usepackage[utf8]{inputenc} 
\usepackage{moreverb}
\usepackage{graphicx}
\usepackage{amsmath}
\usepackage[wide]{sidecap}
\usepackage{lipsum,babel}
\usepackage{setspace}

\begin{document}

\begin{frontmatter}
	
	\title{Design of a Sag and Surge Detector For Residential Voltage}
	\author{Daniel Pérez}\ead{daniel.andres.perez.espina@uru.edu}
	\address{School of Electrical Engineering, Engineering Faculty, Universidad Rafael Urdaneta, Maracaibo, Zulia, Venezuela}
	
	\begin{abstract}
	{An inexpensive protection device capable of disconnecting a load upon detecting a voltage sag or surge is presented. Said device is aimed towards the residential voltage of Venezuela's electrical system. The device is based on the Arduino platform, and uses an ATmega 328P. It detects the presence of a voltage sag or surge by sampling the residential voltage used to feed the load, and continuously measuring the RMS $\frac{1}{2}$ voltage. It disconnects the load after 3 semi-cycles where the residential voltage is outside the boundaries established by Venezuela's national electrical code, and therefore there's the presence of a voltage sag or surge. The device is capable of transmitting the data of the sampled signal, for making further analysis.}
	\end{abstract}
\end{frontmatter}

\section{Introduction}
In present day Venezuela various power disturbances such as voltage sags and surges are very prominent. Even though there isn't any official data disclosed on the topic, it is agreed among many Venezuelans that one of the main sources of damage of electric appliances are volt-age sags and surges. These disturbances arise because Venezuela's electrical system is in a bad state, which is due to mismanagement. The only permanent solution to this issue would be to make the necessary improvements to the electrical system in order to restore its proper functioning.\\
For this reason it can be asserted that any permanent solution to the main power quality issues of Venezuela is mid-term, at least. For as long as the improvements aren't done or even started to do, the voltage sags and surges are going to keep reappearing. This problem gene-rates the necessity of creating a short-term solution, one that doesn't have to be permanent but that definitely has to have swift means of implementation. The solution has to be very low cost as well, due to the precarious si-tuation of the Venezuelan economy.\\
The design of a voltage sag and surge detector based on the Arduino platform is presented as a possible solution. Said device disconnects any load attached to it, whenever the RMS $\frac{1}{2}$ value of the residential voltage signal is outside the boundaries established by Venezuela's national electrical code (NEC) \cite{CEN}, for a period of time equal or higher than 3 semi-cycles. The boundaries set by Venezuela's NEC coincide with the ones established by IEEE \cite{1159}. Nonetheless, being that in Venezuela the residential systems can either use 110 or 120 VRMS, it was decided to use 110 VRMS to set the lower boundary and 120 VRMS to set the upper boundary.\\
To do the detection, the device samples the residential voltage signal at a rate of 3600 Hz, using the code of a power quality monitor developed to address Venezuela's power quality issues as well \cite{dape}. After establishing the presence of either a sag or surge, the device disconnects the load attached to it by triggering a relay connected in series with the load. The device doesn't reconnect the load until the residential voltage signal is in compliance with the set boundaries. The device is capable of transmitting the sampled data it retrieves, which can be used for either a deep analysis after the fact, or real-time mo-nitoring.\\
The device distinguishes itself from other protection devices using the term ``detector'', because usually protectors that are marketed as voltage sag or surge protectors offer some sort of relief upon the disturbance, such as peak suppression, instead of just disconnecting the load.

\section{Hardware}
\begin{figure}[!ht]
\centering
\includegraphics[width=85mm]{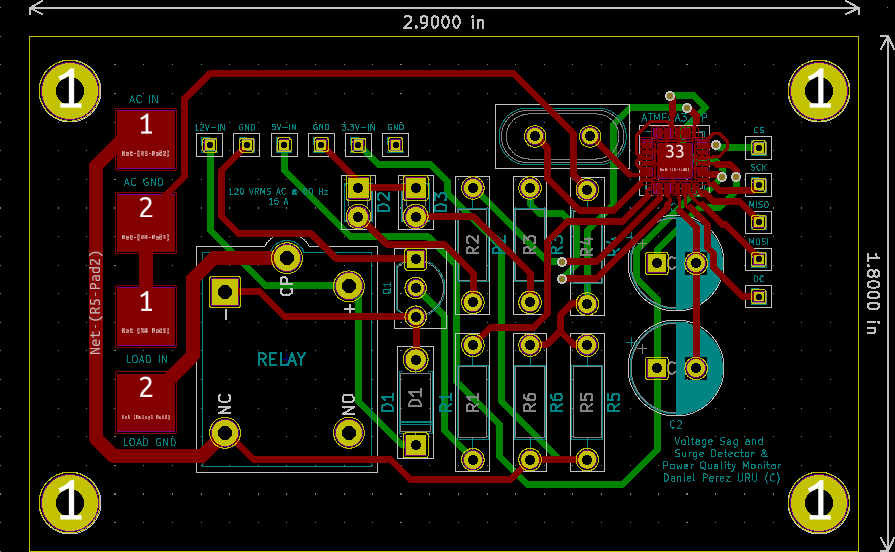}
\caption{Preview of PCB design that encompasses entire design. Doesn't include DC sources}
\label{pcbatmega}
\end{figure}

The hardware used for this device is comprised by a small and simple circuit used for sampling the residential voltage signal, a circuit for managing the load attached to the device and the ATmega328P that controls the entire system.\\
In figure \ref{pcbatmega} can be seen the PCB design of the entire system. It is shown directly instead of presenting the schematic first to maintain simplicity in the presentation, and because it lets showcase the positioning of the elements of the system.
\\For this design it was assumed that separate modules are going to be used to provide the DC voltages required by the system. Being that this design doesn't include a model for the case, this gives total freedom in terms of making a design for it. Also, it is worth noting that each pin used for the serial port interface (SPI) is connected to a pad, in such way that is easy to solder any module that can be used for transmitting the sampled data, which opens up the possibility of assembling an Internet of Things infrastructure \cite{ARD} with this design. If a component to send the data isn't to be added, then the pads of the pins of the SPI should be shorted to ground to avoid leaving them floating.\\
To lower the amount of noise in the system as much as possible, a series of measures taken in the power qua-lity monitor \cite{dape} were employed here as well. Those measures were taken because floating pins generate noise and can trigger $\frac{di}{dt}$ transients in the system \cite{CDC}, and are based on various concepts of transmission lines\cite{MOT}, grounding\cite{GND}, power electronics\cite{VR} and  electromagnetic compatibility\cite{EMC},\cite{EMG}. It should be pointed out that in order to reduce the overall inductance of the system tracks were kept as short as possible, and to increase the overall capacitance the gap between power lines was kept as short as possible.\\
The circuit utilized for sampling is the same used for a power quality monitor targeted for Venezuela's residential voltage \cite{dape}. Said circuit transforms a 120 VRMS at 60 Hz sine wave into one of 0.6 VRMS at 60 Hz with an offset of 3.3 V. This measure was taken to stay in compliance with the limits of the analog to digital converter (ADC) of the ATmega 328P \cite{AVR}. This design was preferred because using a transformer would unnecessarily add noise to the system, and it would incorporate losses as well. The use of resistors in this scenario is valid because the measurement of an analog signal is done with a shunt connection, and the value of the internal resistance of the ADC of the ATmega 328P is 100M$\Omega$, which is 20000 times higher than the value of the resistor that perceives the input analog signal.\\
To manage the load, a relay with a normally open switch is connected in series with it. The ATmega 328P only closes the switch when the code determines that proper conditions for powering the load are met, according to the set boundaries. As can be seen in figure \ref{pcbatmega}, there's a diode connected to the coil of the relay, and it has the function to prevent that a negative voltage is seen by the pins of the ATmega 328p, scenario that can happen due to a $\frac{di}{dt}$ transient.\\
A couple of LEDs are used to indicate if the residential voltage signal is currently meeting the requirements. The red LED indicates that the voltage is beyond the 10\% above the 120 VRMS or 10\% under 110 VRMS, while the green LED points that the system is providing adequate energy. Adding a small display to indicate the current stats of the residential voltage signal is planned to be added in the future.
An alternative design using an Arduino NANO V3 is shown in figure \ref{pcbnanov3}:
\begin{figure}[!ht]
\centering
\includegraphics[width=85mm]{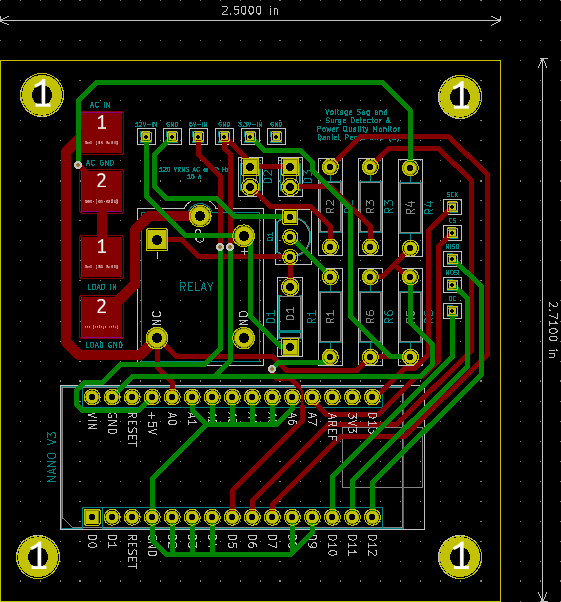}
\caption{Preview of an alternative PCB design that uses an Arduino NANO V3 board}
\label{pcbnanov3}
\end{figure}
\\This alternative version works exactly the same as the original one. The main difference lies in the fact that updating the software is a simple and straightforward process in this alternative design, because it only requires using a PC with the Arduino IDE and a mini USB cable. Also, it is capable of transmitting data only using the mini USB cable if desired.
\section{Software}
Although this device can be connected to others in order to send data, its only function beyond providing output is to manage the connection of the load attached to it, based on the current status of the residential voltage signal it is sampling. Therefore, all the code that it requires to do these tasks has to be run by the ATmega 328P. For this version, all the code used is based on the Arduino platform's set of C/C++ instructions. This requires flashing to the ATmega 328P the bootloader of Arduino, and the program to be used.
There's two core mechanics present in the software of this device: the data sampling and data processing.\\
\subsection{Data Sampling}
It is based on the code used for sampling in a power quality monitor targeted for Venezuela's residential voltage \cite{dape}. The code ensures that sampling occurs at 3600 Hz, which is well beyond the Nyquist frequency for this scenario \cite{DSP}, which would be 120 Hz. It is worth noting that although a much higher sampling rate isn't required, the more values taken from the residential voltage, more precise the RMS $\frac{1}{2}$ value will be. The code used for this device is planned to be updated in the future, being that the code it's based on is subject to future updates as well. This is because it is desired to ensure compatibility between this device and the power quality monitor.\\
\subsection{Data Processing}
The data processing in this device is fairly simple, it only requires to recognize when the semi-cycles begin, and to adjust the data accordingly to calculate the RMS $\frac{1}{2}$ value of each semi-cycle. In addition to this, the code that corresponds to the data processing also keeps track of the overall behavior of the RMS $\frac{1}{2}$ values.\\
The expression used to calculate the RMS $\frac{1}{2}$ value for each semi-cycle, is the following:
\begin{equation}
  VRMS\frac{1}{2} = \frac{\sqrt{\sum_{n=1}^{60}V_{n}^{2}}}{60}
\end{equation}
Where $V_{n}$ corresponds to each sampled voltage, and 60 represents the amount of samples taken per each semi-cycle.\\
The code determines the presence of either a voltage sag or surge after the RMS $\frac{1}{2}$ value meets the criteria for either of these disturbances, for 3 consecutive semi-cycles. Once the presence is established the code orders the disconnection of the load, and it doesn't get reconnected until the RMS $\frac{1}{2}$ values are in compliance with the set boundaries, for 360 consecutive semi-cycles. This way of operation was chosen based on the fact that all types of loads don't suffer any consequences from sags with a duration under 5 cycles \cite{PSERC}, and because most voltage surges that appear in Venezuela's electrical system come after voltage sags, as a consequence of the FIDVR phenomenon \cite{FIDVR}. This ensures that the impact of voltage sags and surges to the load is kept at a minimum, but it also implies that this device is merely a palliative. The abrupt disconnection of an appliance isn't a harm-free operation, it is just the most desirable option upon a voltage sag or surge under the current scenario that Venezuela is in.\\
A graphical representation of the algorithm that governs the sag and surge detection and load management can be seen in figure \ref{alg}:
\begin{figure}[!ht]
\centering
\includegraphics[width=85mm]{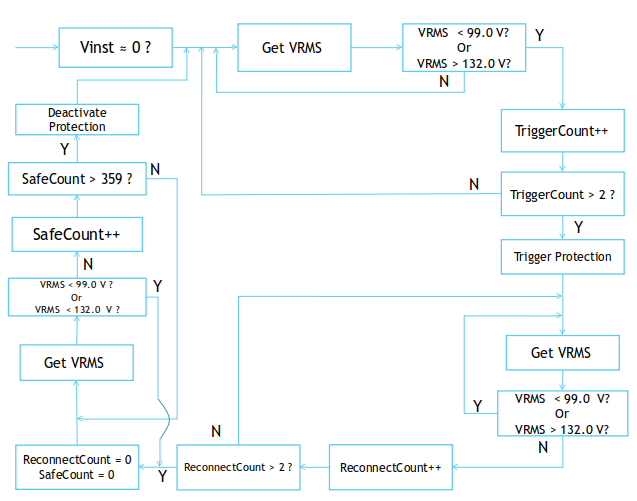}
\caption{Algorithm used for detecting voltage sags and surges, and sensing proper conditions for reconnecting the load}
\label{alg}
\end{figure}

\section{Conclusions}
The design requires the addition of a case in order to be ready for commercial use, although the core mechanics are ensured to work with the current version. To this design can be added any component that can be used for data transmission, which opens up the possibility of building an Internet of Things infrastructure based on it \cite{ARD}. It can be used even as a sensor for a micro-grid system, due to the scope it can reach. The design is very simple and uses low cost parts.

\end{document}